# On the Kinetic Theory of Subauroral Arcs

## Evgeny Mishin[1] and Anatoly Streltsov[2]

[1]Air Force Research Laboratory Space Vehicles Directorate, Albuquerque, NM, USA.

[2]Department of Physical Sciences, Embry-Riddle Aeronautical University, Daytona Beach, FL, USA.

Corresponding author: Evgeny Mishin (evgeny.mishin@spaceforce.mil)

**Key Points:**

Ionospheric feedback makes strong small-scale field-aligned currents and electric fields in fast subauroral flows with low-density troughs.

Solving the kinetic equation with these fields gives suprathermal electrons and excited neutrals explaining the subauroral arcs features.

The developed theory predicts that some subauroral arcs might have the transient phase with typical aurora-like emissions fading afterwards.

### Abstract

We report on numerical and theoretical investigation of subauroral arcs within strong subauroral ion drifts (SAID)--STEVE and Picket Fence. Their explanation requires the specific suprathermal electron distribution and excitation of vibrational and electronic states of neutral species. We show that the ionospheric feedback in the strong SAID with low-density troughs with or without a so-called ionospheric valley generates intense, small-scale field-aligned currents and electric fields below the $F_2$ peak. With these fields, we solved the Boltzmann kinetic equation for the distribution of electrons and power for excitation and ionization of neutral gas (the energy balance). The obtained suprathermal electron population and energy balance are just what is necessary for Picket Fence. Concerning STEVE, the kinetic theory predictions are in a good qualitative agreement with its basic features. Besides, the theory predicts that subauroral arcs might have the transient phase with typical aurora-like emissions that fade out afterwards.

### Plain Language Summary

Subauroral arcs inside fast convection flows with low-density troughs and elevated electron temperature are termed STEVE and Picket Fence. Their optical spectra are radically different from usual aurora and Stable Auroral Red (SAR) arcs and require the specific electron distribution and excitation of vibrational and electronic states of neutral species. Our modeling of the feedback instability in the low-density, high-velocity flow channels reveals greatly enhanced parallel electric fields developing in concert with small-scale field-aligned currents. Solving the Boltzmann kinetic equation with these fields, we found the electron distribution and the power going to excitation and ionization of neutral gas (the energy balance). The obtained suprathermal electron population and excitation of neutral species are just what is required for Picket Fence. The kinetic theory predictions are in a qualitative agreement with STEVE's basic features. The theory predicts that subauroral arcs might have the transient phase with typical aurora-like emissions that fade out with time.

## 1 Introduction

East-west-aligned, mauve "ribbons" equatorward of the auroral zone have long been known under a quaint name "Steve". Those are related to subauroral ion drifts (SAID) with the westward speeds of $|\mathbf{v}_W| = |\mathbf{E}_\perp \times \mathbf{b}_0| \sim 4-6$ km/s ($\mathbf{b}_0 = \mathbf{B}_0 / B_0$), elevated electron temperatures, $T_e$, up to $\sim 10^4$ K, and deep density troughs, $n_e \leq 10^4$ cm$^{-3}$ (e.g., MacDonald et al., 2018; Archer et al., 2019a). Due to large speeds and strongly elevated $T_e$, the Steve arcs were dubbed Strong Thermal Emission Velocity Enhancement (STEVE). An extensive account of earlier works prior to the "STEVE era" is given in (Hunnekuhl & MacDonald, 2020; Henderson, 2021). STEVE is the premidnight, substorm recovery phenomenon (Gallardo-Lacourt et al., 2018a) lacking >50 eV electrons (Gallardo-Lacourt et al., 2018b; Chu et al., 2019; Nishimura et al., 2019). STEVE's unusual color is determined by a 400-800 nm continuum (Gillies et al. 2019).

Sometimes, STEVE is accompanied by a greenish-rayed arc resembling a picket fence, hence termed Picket Fence. Liang et al. (2019) reported on a subauroral arc with two emission structures at altitudes $h \leq 150$ km and ~250 km. The latter contained substantial enhancement of the redline emission from the O($^1$D) state over background. Archer et al. (2019b) estimated the height extent of STEVE and Picket Fence on nearby or perhaps the same magnetic field lines as 130-270 and 97-150 km, respectively.

During a STEVE-Picket Fence event captured in the northern hemisphere (NH), Nishimura et al. (2019) used DMSP F17, THEMIS-E (TH-E), and Swarm A in the southern hemisphere (SH) close to the NH track of Swarm-B. The Swarm satellites detected similar SAID channels, typical of the SAID events, including particle distributions near the magnetic equator (e.g., Mishin, 2013). The F17 and TH-E electric fields exhibited a double-SAID structure. The outermost SAID was magnetically conjugate to Picket Fence and nearly collocated with enhanced energetic electron population ("bump"), ~10 keV, in the plasmasphere and top ionosphere. The innermost SAID channel without energetic electrons was conjugate to STEVE. Thus, Nishimura et al. (2019) concluded that Picket Fence is, in fact, a rayed subauroral aurora.

Mishin and Streltsov (2019; henceforth, MS19) have shown that the presence of the energetic "bump", which enhances the Hall conductance ($\Sigma_H$) over the Pedersen conductance ($\Sigma_P$), leads to the Picket Fence formation. MS19 employed a three-dimensional (3D), model (Jia & Streltsov, 2014) of the ionospheric feedback instability (IFI) in the SAID channel. In a 2D system, without the Hall current, east-west-aligned "sheets" of small-scale upward and downward FACs carried by dispersive Alfvén waves are closed by the meridional Pedersen current. The characteristic scale of the resulting series of east-west-aligned strips is of the order of the most unstable wavelength. The Hall conductance ($\Sigma_H > \Sigma_P$) makes a 3D system by rotating the IFI-generated ionospheric currents and electric fields, which results in a chain of small-scale vortices resembling a series of "pickets" within the SAID channel (e.g., MS19, Figure 3).

On the other hand, auroral emissions produced by collisional degradation of energetic electrons exhibit the well-defined spectrum. It is formed mainly by the so-called degradation spectrum of suprathermal electrons (e.g., Banks et al., 1974; Konovalov & Son, 2015):

$$F_s(\varepsilon) = \frac{m_e^2}{2\varepsilon}\Phi_e(\varepsilon) \approx \frac{3}{2\pi v_c^3}n_s \begin{cases} (\varepsilon_c/\varepsilon)^{4.5} & \text{at } \varepsilon_c \leq \varepsilon \leq \varepsilon_c^* \\ (\varepsilon_c/\varepsilon_c^*)^{4.5}(\varepsilon_c^*/\varepsilon)^{3.5} & \text{at } \varepsilon_c^* < \varepsilon < 300 \text{ eV} \end{cases} \quad (1)$$

Here $v_c = \sqrt{2\varepsilon_c/m_e}$, $\varepsilon_c \approx 5$ eV, $\varepsilon_c^* \approx 20$ eV, $n_s \approx (25-30)n_b$, $n_b$ is the density of precipitating energetic electrons with the omnidirectional differential number flux $\Phi_e(\varepsilon)$.

However, in the Picket Fence events (Mende et al., 2019; Mende & Turner, 2019) the blue-line emission at 427.8 nm from the $N_2^+1N(0,1)$ level was lacking, while the first positive band ($N_21P$) ≥650-nm emissions via $B^3\Pi_g \rightarrow A^3\Sigma_u^+ + h\nu$ transition and the green line from O($^1$S) were abundant. That is, the suprathermal population significantly increased over $F_s(\varepsilon)$ (1) between $\sim\varepsilon_c$ and the $N_2^+1N$ threshold, $\varepsilon_b = 18.75$ eV. Note, only ≈2.3% of the $N_2$ ionization radiates the blue line. An additional source of the green line is collisional quenching of the metastable $N_2(A^3\Sigma_u^+)$ state by atomic oxygen leading to energy transfer to O($^1$S). As the metastable O($^1$D) state is strongly quenched below ~170 km, this reaction leads mainly to the green color.

Furthermore, contrary to the thermal excitation, the O($^1$D) redline emission in STEVE is on average less intense than in SAR arcs with significantly smaller temperatures. Using Mishin et al.'s (2000; 2004) kinetic solution, MS19 (Figure 4) demonstrated that the $N_2$ "vibrational barrier" practically eliminates the thermal excitation in STEVE because of the electron distribution function (EDF) "bite-out". The term "vibrational; barrier" designates a greatly enhanced cross-section of $N_2$ vibrational excitations, $\sigma_V(\varepsilon)$, in the energy range $\varepsilon > \varepsilon_1 = \sqrt{m_e v_1^2} \approx 1.9$ eV and $\varepsilon \leq \varepsilon_2 \approx 3.5$ eV. Based on this electron kinetic effect, MS19 argued that excitation by the suprathermal electron population determines the STEVE continuum.

Therefore, the question arises about the source of suprathermal electrons at the STEVE and Picket Fence altitudes. It is known that suprathermal electrons, $\varepsilon \sim 10-300$ eV, in the SAID channel in the top ionosphere come from the turbulent plasmasphere boundary layer (e.g., Mishin, 2013). However, this population degrades along the path to the F region, not to mention the E region (see Khazanov et al., 2017). The obvious corollary to the above is that an unknown local source of low-energy, $\varepsilon < 18.75$ eV, suprathermal electrons and $N_2$ excitation operates in the SAID channel below ~270 km. Mishin and Streltsov (2021, Chapter 5.3; hereafter MS21, Ch.5.3) suggested to invoke parallel electric fields, $E_\parallel$, resulting from an instability driven by intense FACs. This is most likely (Voronkov & Mishin, 1993) in a plasma density depletion (a so-called "valley") of $n_e^{(v)} \sim 10^3$ cm$^{-3}$ between ≥120 and ≤200 km (Titheridge, 2003). In addition, $E_\parallel$ is the intrinsic feature of the IFI- generated small-scale dispersive Alfvén waves.

This paper investigates the suprathermal electron population produced by small-scale parallel electric fields generated by the IFI inside a strong westward flow channel with a deep density trough. First, we numerically simulate the IFI development with the input parameters, such as the driving poleward electric field and the electron density, similar to those in the STEVE region in the top ionosphere. The density altitude profile in the STEVE region has not yet been determined, especially below the F$_2$ peak. Thus, we assume an arbitrary profile based on the nighttime values (e.g., Titheridge, 2003) and do not attempt to address specific experimental details but rather focus on the basic features of the simulation. In any case, as noted by MS19, the equilibrium density profile in the SAID/STEVE region would be modified by the upwelling in the atmosphere due to enhanced ohmic heating.

We show that the IFI driven by strong electric fields within a low-density trough leads to greatly enhanced $E_\parallel$ and the parallel voltage below the nighttime F$_2$ peak. The presence of the valley further increases $E_\parallel$ and the voltage. The obtained electric fields are used as the input into the Boltzmann kinetic equation to find the EDF and the power going to excitation and ionization of neutral species. The simulation results show the feasibility of this mechanism for subauroral arcs. In particular, it creates the population of suprathermal electrons and $N_2$ excitation in good quantitative agreement with that required for Picket Fence. As far as the STEVE spectrum is concerned, the kinetic theory predictions are in a qualitative agreement with its basic features.

## 2 Ionospheric Feedback Instability in Strong SAID

We calculate the parallel electric field created by the IFI in a fast SAID channel with a deep density trough. The ionospheric feedback process amplifies small-scale Alfvén waves by virtue of over reflection from the ionosphere, which is caused by the shear convection flow due to the altitudinal dependence of ion-neutral collisions. In particular, for E-region plasma densities, $n_{0E} \leq 10^4$ cm$^{-3}$, the IFI threshold for 1-10 km wavelengths is approximately $E_{th} \approx 50\Omega_{ci}/\nu_{i0}$ mV/m (Trakhtengerts & Feldstein, 1991). Here, $\nu_{i0}$ ($\Omega_{ci}$) is the ion-neutral collision (ion cyclotron) frequency at 105-110 km. In a 2D model, without the Hall current, the Pedersen (meridional) current, $\mathbf{J}_P = \Sigma_P \mathbf{E}_\perp$, has the form of a strip between the east-west aligned sheets of upward and downward FACs enclosing the flow channel. The IFI "splits" the initial strip into a series of small-scale strips determined by the most unstable wavelength. The resulting fine meridional structure of the FACs inside the STEVE channel is consistent with the Swarm-A FACs in the Nishimura et al., 2019 (Figure 2j) event.

A two-fluid MHD model (e.g., Streltsov et al., 2012; Streltsov & Mishin, 2018; hereafter SM18) consists of the "magnetospheric" and "ionospheric" parts. The former describes dispersive Alfven waves in an axisymmetric dipole magnetic field using equations for the electron parallel momentum and continuity of the plasma density, $n$, and FACs, $j_\parallel = -n_e e u$. Here $u$ is the parallel component of the electron velocity.

We take a meridional (poleward) trapezoidal electric field, $E_x(x)$, centered at $L = 4.9$ and consider a 2D problem justified for azimuthally extended STEVE arcs. As the amplitude of the IFI-generated small-scale FACs is larger in the 3D case (e.g., MS19), it is anticipated that simulations in a full 3D geometry for the same input conditions would give a larger magnitude of $E_\parallel$. As in SM18, the computational domain represents a 2D slice of the axisymmetric dipole magnetic field between magnetic shells 4.7 and 5.1. The ionospheric boundaries of the domain are set at 110-km altitude. The vertical size of the conducting portion of the E-region ionosphere is much less than the parallel wavelength, so it is taken as a narrow layer with the uniform density and electric field. In this case, the simplest (so-called electrostatic) boundary conditions are derived by integrating the current continuity equation, $\nabla \cdot \mathbf{j} = 0$, over the conducting (dynamo) layer with the effective thickness of $h \approx 10-20 \, \text{km}$

$$\nabla \cdot \left( \Sigma_P \mathbf{E}_\perp \right) = \pm j_{\parallel, i} \tag{2}$$

Here $j_{\parallel, i}$ is the FAC density on the top of the E region and the sign "+/-" in the right-hand side of equation (5) is for the southern/northern hemisphere. The variation of the E-region density is derived by integrating the density continuity equation over the conducting layer

$$\frac{\partial n}{\partial t} + \nabla_\perp \cdot \left( n \mathbf{v}_E \right) = \frac{j_\parallel}{eh} + \alpha \left( n_0^2 - n^2 \right) \tag{3}$$

Here $\mathbf{v}_E$ is the electric drift velocity; $\alpha$ is the recombination coefficient; the term $\alpha n^2$ represents losses due to the recombination, and $\alpha n_0^2$ represents all unspecified sources of the ionospheric plasma that provide the equilibrium state of the ionosphere $n_0$. This system of equations forms a positive feedback loop: $\delta\Sigma_P(\delta n_e) \rightarrow \delta j_\parallel \rightarrow \delta n_e$ (see, e.g., SM18 for details).

The numerical procedure used to solve the model equations has been described in detail in several papers (e.g., Streltsov et al., 2012; SM18; and references therein). Here, we present only the simulation results. As in the case considered in SM18, we assume asymmetric plasma profiles in the hemispheres. The northern hemisphere has no valley and the plasma density between 110 km (E region) and 300 km (F$_2$ peak) increases from $n_E^{(N)} = 10^4$ to $n_F = 9 \times 10^4$ cm$^{-3}$. In the runs when the southern hemisphere has no valley, the density at 110 km is taken $n_E^{(S)} = 3 \times 10^3$ cm$^{-3}$. In the runs with the SH valley I, we assume that $n_e^{(v)} = \text{const} = n_E^{(S)}$ between 110 and 200 km and increases to $n_F$ between 200 and 300 km. In the case of the SH valley II, the E-region density at ≤110 km remains $n_E^{(S)}$ but between ≈110 km and 200 km $n_e = n_E^{(S)}/2$; increasing to $n_F$ thereafter. The background plasma temperature is taken 1000 K.

Figure 1 illustrates the simulation results for $\max(E_X(x)) \equiv E_0 = 100$ mV/m with the 2D spatial distribution in the computational domain shown in top frame. As typical (e.g., SM18), the IFI creates in both hemispheres a system of intense small-scale FACs, with downward currents ($j_\downarrow$) dominating their upward ($j_\uparrow$) counterparts. The current intensities of the order of 50-80 µA/m$^2$ are comparable to those observed in the auroral ionosphere (e.g., Akbari et al., 2022).

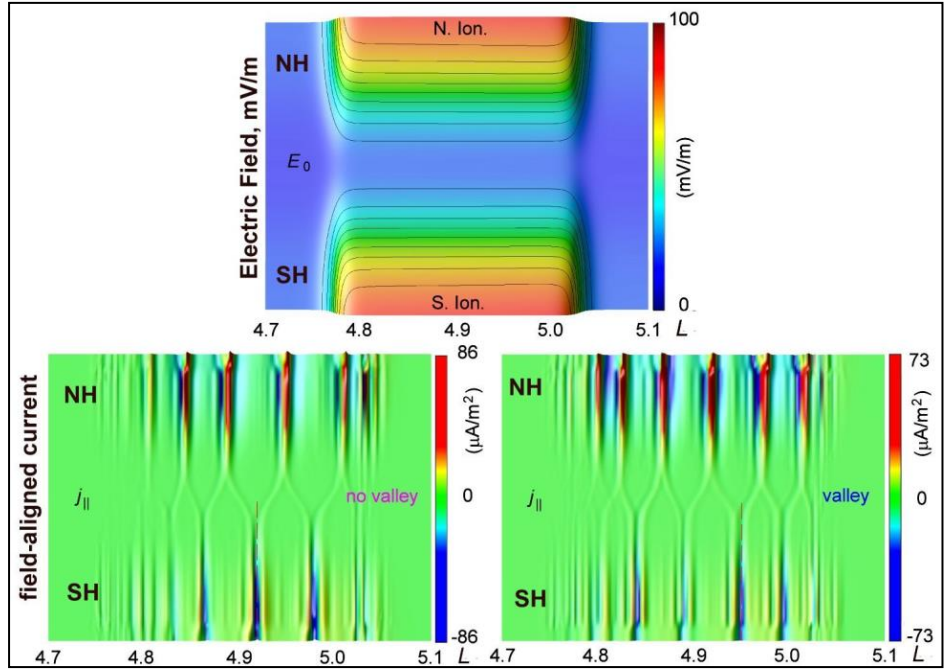

**Figure 1**. IFI simulation for $E_0 = 100$ mV/m: The spatial distribution in the computational domain of (top) the driving electric field and field-aligned currents (left) without and (right) with the southern hemisphere(SH) valley I (see text).

It is not surprising that such intense currents are associated with enhanced parallel fields, creating a 0.5-1 kV voltage, $|\phi_\||$, between 110 and 200 km, as shown in Figure 2. The resulting power consumption due to Joule heating (insets) amounts to $Q_J = j_\| E_\| = 100 - 900$ nW/m$^3$. The field magnitude, $|E_\||$, and the voltage increase nonlinearly with the driving field, $E_0$. Overall, in the lower-density southern hemisphere the field magnitude, $|E_\|^{(SH)}|$, and voltage, $|\phi_\|^{(SH)}|$, are larger than their northern counterparts. On the contrary, the FAC intensities, $|j_\|^{(SH)}|$, are smaller than $|j_\|^{(NH)}|$. For example, for $E_0 = 100$ mV/m without (with) the valley we have $E_\|^{(SH)} / E_\|^{(NH)} \approx 2$ ($\approx 1.5$) and $j_\|^{(SH)} / j_\|^{(NH)} \approx 0.7$ ($\approx 0.23$). Taking the symmetric hemispheres without the valley results in about the same amplitudes as in Figure 2c without the valley but quite low electric fields and currents in the runs with the valley I and II.

Hence, the IFI driven by strong electric fields in a low-density channel leads to greatly enhanced parallel electric fields and Joule heating below the nighttime F$_2$ peak. Their effect on excitation of neutral gas is described next.

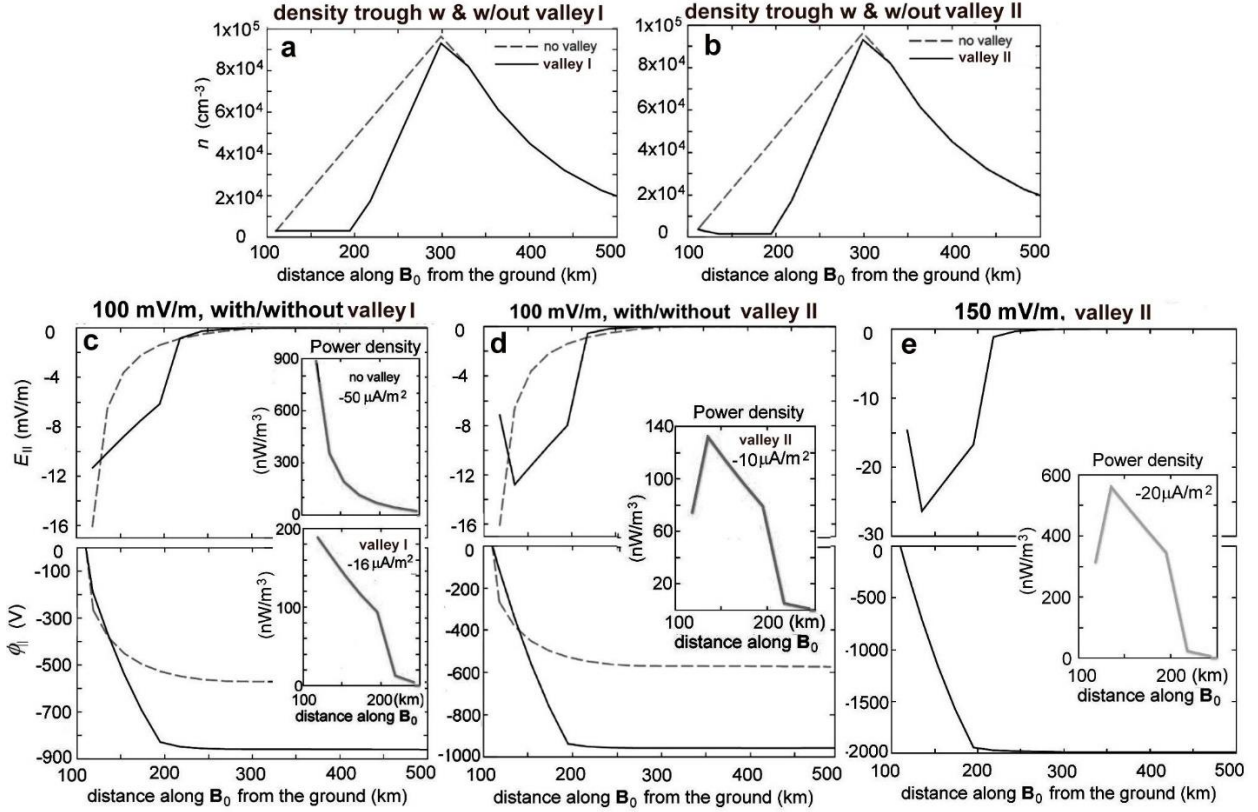

**Figure 2.** (Top) Variation along the magnetic field line of the plasma density in the trough without (the dashed line) and with (solid) the valley (a) I or (b) II; (mid) $E_\parallel$ and (bottom) $\Phi_\parallel$ taken in the region of the most intense SH downward current without (dashed lines) and with (solid) the valley I and II for $E_0$ = 100 mV/m and 150 mV/m. Insets: Variation of the corresponding power density, $Q_J = j_\parallel E_\parallel$ in nW/m³; the FAC density averaged over the distance along $\mathbf{B}_0$ between 100 and 200 km, $\langle j_\parallel \rangle$, is indicated.

### 3 Electron Distribution and the Energy Balance

The power distribution over inelastic processes (the energy balance) is calculated via the electron distribution function (EDF), $F_e(\varepsilon)$, as follows

$$P_j = \varepsilon_j \int_{\varepsilon_j}^{\infty} \nu_j(\varepsilon) F_e(\varepsilon) d^3v = 4\pi \varepsilon_j N_j \int_{\varepsilon_j}^{\infty} \sigma_j(\varepsilon) \Phi_e(\varepsilon) d\varepsilon \qquad (4)$$

Here $\nu_j(\varepsilon) = v\sigma_j(\varepsilon)N_j$ is the collision frequency of an inelastic process with the cross-section $\sigma_j$ and the excitation energy $\varepsilon_j$, as exemplified in Figure 3a. The EDF under action of the imposed electric field is a solution of the Boltzmann kinetic equation that includes elastic and inelastic processes (e.g., Capitelli et al., 2000). Figure 3b shows a rigorous numerical solution (Milikh & Dimant, 2003), which exemplifies the EDF bite-out in the E region. Figure 3c shows the energy balance explicitly calculated by Dyatko et al. (1989) with the composition typical of 110-120 km (e.g., Picone et al., 2002) for the input values of $E_\parallel / N_n \equiv \tilde{E}$ in the Townsend units: 1 [Td] = $10^{-17}$ [V·cm²]. Here, the neutral density, $N_n$, is the sum of the densities of $N_2$ ($N_{N2}$), $O$ ($N_O$), and

$O_2$ ($N_{O2}$). For reference, for $N_n = 10^{12}$ cm$^{-3}$ at 120 km, $\tilde{E} = 1$ Td yields $E_{\parallel} = 10^{-17} N_n$ [V/cm] = 1 [mV/m].

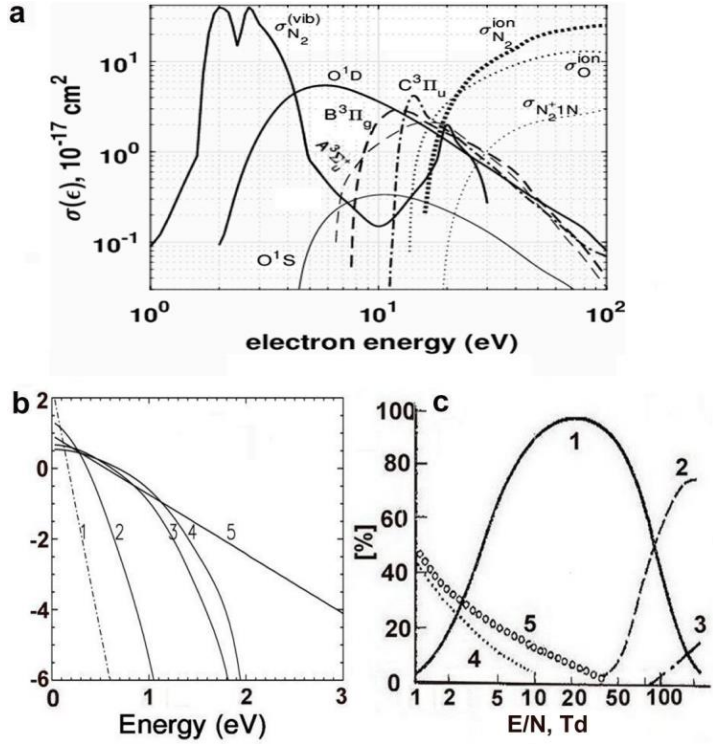

**Figure 3**. (a) Some of the basic ionization and excitation cross sections of $N_2$ and $O$ calculated with the Majeed and Strickland (1997) and Itikawa (2006) data. (b) Electron distribution function versus energy at $N_{N2}$: $N_{O2}$: $N_O$ = 75:10:15 ($h \approx 110$ km): 1 & 5- Maxwellian at $T_e$ = 360 & 3000 K; 2- $E_{\parallel}$ = 1.1 mV/m ($T_e$ = 1290K); 3- $E_{\parallel}$ = 2.1 mV/m ($T_e$ = 2970K); 4- $E_{\parallel}$ = 4.3 mV/m ($T_e$ = 3530K). Adapted from Milikh & Dimant (2003). (c) Energy distribution over various channels versus $E_{\parallel}/N_n$ at $N_{N2}$: $N_{O2}$: $N_O$ = 77:8:15 ($h \approx 120$ km): 1- vibrational levels of $N_2$, 2 – electronic levels of $N_2$, 3 – ionization of $N_2$, 4 – vibrational levels of $O_2$, 5 – gas heating. From Dyatko et al. (1989). It is assumed that the initial vibrational and fine structure populations have the gas temperature, $T$ = 550 K, while the initial electronic levels are void.

The power lost in vibrationally excited $N_2$ ($P_V$) exceeds 50% at $\tilde{E} > \tilde{E}_V \approx 5$ Td and reaches a broad maximum ~90% at $\tilde{E}_V^{(max)} \sim 20$ Td. Excitation of the $N_2$ electronic levels at $\tilde{E} = \tilde{E}_A \approx 40$ and $\tilde{E}_\Sigma \approx 100$ Td takes away the power, $P_A \approx 3\%$ (mainly the $A^3\Sigma_u^+$ state) and $P_\Sigma \approx 50\%$, respectively. Of $P_\Sigma \approx 50\%$, about 30% goes into the triplet $A^3\Sigma_u^+$ ("A"), $B^3\Pi_g$ ("B"), and $C^3\Pi_u$ ("C"), with the excitation energies $\varepsilon_A$ = 6.17 eV, $\varepsilon_B$ = 7.35 eV, and $\varepsilon_C$ = 11.03 eV, respectively. At the same time, about 10% goes to the electronic levels of $O_2$ and $O$ (not shown), mainly O($^1$D) with $\varepsilon_r$ = 1.96 eV and O($^1$S) with $\varepsilon_g$ = 4.17 eV. Note that the ionization of $N_2$ (curve 3) begins at $\tilde{E} > 90$ Td and amounts to $P_{ion}$ ~2% at 100 Td. That is, the ionizing population, $\varepsilon > \varepsilon_{ion}$ = 15.6 eV, appears at $\tilde{E} > \tilde{E}_{ion} \approx 90$ Td.

Note, the suprathermal tail at $\tilde{E}_A \leq \tilde{E} \leq \tilde{E}_{ion}$ is confined within the energy range optimal for Picket Fence, $\varepsilon_2 < \varepsilon \leq \varepsilon_{ion}$, and a significant power is spent on the $N_2$ excitation, as well. At $h$ =

120 km, the above values of $\tilde{E}_V, \tilde{E}_A$, and $\tilde{E}_{ion}$ correspond to the magnitudes, $E_V = 5$, $E_A = 40$, and $E_{ion} = 90$ mV/m, respectively, and thus the IFI generated fields in Figure 2 at $E_0 \leq 150$ mV/m excite mainly $N_2$ vibrational states. If the energy balance at higher altitudes remains the same as at 120 km, then the corresponding values will reduce as $N_n(h)/10^{12}$. Particularly, for $N_n = 10^{11}$ ($10^{10}$) cm$^{-3}$, i.e., $h \approx 135$ (175) km, we get $E_V = 0.5$ (0.05), $E_A = 4$ (0.4), and $E_{ion} = 9$ (0.9) mV/m. Therefore, as one can see from Figure 2, the generated fields in the trough exceed $E_{ion}$ at and above 135 km with the valley and at $\geq 175$ km without the valley.

Most notably, for $E_0 = 100$ mV/m in the density trough without the valley, we get at $h = 135$ km $E_A < E_\parallel \approx \frac{1}{2}E_{ion}$, with $P_V \leq 80\%$ and $P_\Sigma \geq 10\%$. Further, the neutral density usually obeys the barometric law and hence decreases with altitude faster than $|E_\parallel|$ in Figure 2. Accordingly, the Picket Fence-required conditions, i.e., the enhanced population between $\sim\varepsilon_c$ and $\varepsilon_b$ and the excited $N_2$ triplet, will be formed at altitudes $135 < h_{pf} < 145$ km. Here, collisional quenching of the metastable O($^1$D) state severely suppresses the redline emission (e.g., Mishin et al., 2004, Figure 5). Therefore, the green-line and $N_2$1P emissions will make the Picket Fence-like color in accordance with the observations (Mende et al., 2019; Mende & Turner, 2019). Yet, we have to ensure that the EDF and the corresponding energy balance at higher altitudes remain close to those at 120 km. This assertion is addressed next.

Strictly speaking, the EDF and the energy balance at any given altitude should be calculated with the pertinent neutral composition. Such a formidable task is beyond the scope of this paper aimed at the basic qualitative outcomes of the IFI-generated electric fields in a given density profile. Nonetheless, we can make ballpark estimates with the aid of an analytic solution of the Boltzmann kinetic equation. In the $N_2$ vibrational barrier, $\varepsilon_1 \leq \varepsilon \leq \varepsilon_2$, we have $\varepsilon\sigma_V(\varepsilon)F_t(\varepsilon) \gg (\varepsilon+\varepsilon_V)\sigma_V(\varepsilon+\varepsilon_V)F_t(\varepsilon+\varepsilon_V)$, where the energy quantum $\varepsilon_V \approx 0.29(V+1/2)$ eV. This inequality allows using the discrete losses approximation for the collision integral, $St_{dl}(F_e) \approx -\nu_{il}(\varepsilon)F_e \approx -v\sigma_V(\varepsilon)N_{N2}F_e$, yielding (cf. Gurevich, 1978, eq. (2.154))

$$\frac{\partial}{\partial t}F_t(v) - \frac{\partial}{v^2\partial v}v^2 D_\parallel \frac{\partial}{\partial v}F_t(v) = St_{dl}(F_e) \approx -\nu_{il}(\varepsilon)F_t(v) \tag{5}$$

Here $D_\parallel = \frac{1}{3}\nu_e(\varepsilon)u_\parallel^2(\varepsilon)$ and

$$u_\parallel(\varepsilon) = eE_\parallel/m_e\nu_{en}(\varepsilon) \approx 2.7\times 10^6 \varepsilon^{-1/2}\tilde{E}/(\rho_1\tilde\sigma_1 + \rho_2\tilde\sigma_2) \text{ [cm/s]} \tag{6}$$

Here $\rho_{1(2)} \approx N_{1(2)}/(N_1+N_2)$ is the abundance of molecular nitrogen ("1") and atomic oxygen ("2") and the corresponding transport cross-sections $\sigma_j = 10^{-16}\tilde\sigma_j$ cm$^2$; the energy $\varepsilon$ is in eV. The contribution of molecular oxygen--a minor species at $h > 120$ km-- is neglected, as well as Coulomb collisions. As $\max(\sigma_V) = \sigma_V^{(vb)} \approx 4\times 10^{-16}$ cm$^2$ exceeds the O($^1$D) excitation cross-section, $\sigma_D(3\text{eV}) \equiv \sigma_D^{(vb)}$ by a factor of ~30 (Figure 3a), the vibrational barrier remains the key feature that shapes the EDF at altitudes with $\rho_2/\rho_1 < 30$, i.e., below ~300 km (e.g., Mishin et al., 2000).

Disregarding inelastic collisions yields a stationary solution of equation (5) with the constant flux of heated electrons into the suprathermal tail

$$J_t = v^2 D_\| \frac{\partial}{\partial v} F_t(v) = \text{const}(v) \tag{7}$$

Inelastic losses prevent the tail stretching toward large energies, viz., "bite out" the tail population. This process is usually described considering the kinetic equation as the stationary Schrödinger equation in the velocity space, where inelastic collisions represent a potential barrier. The effective "potential" is $\kappa_v^2 \equiv v_{il}/D_\| = 3v_{il}(\varepsilon)/v_e(\varepsilon)u_\|^2(\varepsilon) = 3\delta_{il}/u_\|^2(\varepsilon)$, where $\delta_{il}(\varepsilon)$ is the coefficient of inelastic losses. The tail expansion slows down when $\kappa_v^2(\varepsilon_1) \gg (2\varepsilon_V/m_e)^{-1} \equiv v_{vb}^{-2}$ or

$$\delta_{il}^{(vb)} \equiv v_{il}(\varepsilon_1)/v_e(\varepsilon_1) \gg \tfrac{1}{6} m_e u_\|^2(\varepsilon_1)/\varepsilon_V \tag{8}$$

Owing to this condition, the kinetic equation (5) can be solved in the quasi-classical approximation (cf. Gurevich, 1978, eq. (2.166))

$$F_t(\varepsilon) \approx \frac{C_t}{v(\kappa_v D_\|)^{1/2}} \exp\left(-\int_{v_1}^{v} \kappa_v(v) dv\right) \tag{9}$$

The constant $C_t$ is defined by the matching condition, $F_t(v_1) = F_0(v_1)$, with the EDF of the bulk electrons at $\varepsilon < \varepsilon_1$. Note that electron-electron collisions become important in the F$_2$ region and the EDF at $\varepsilon > \varepsilon_1$ is described by Mishin et al.'s (2004) solution after swap $e^2 E_\|^2/m_e v_e^2$ for $W_{tot}/n_0$.

Outside the barrier, at $\varepsilon < \varepsilon_1$ and $\varepsilon > \varepsilon_2$ (but $<\varepsilon_{ion}$), the lost energy quanta are small, $\Delta\varepsilon_j \ll \varepsilon$, so the continuous loss approximation is applicable:

$$St_{cl}(F_e) \approx \frac{1}{2v^2} \frac{\partial}{\partial v}\left(v^3 L_{il}(v) F_e\right)$$

Here $L_{il}(v) = \sum_j v_j(v)\varepsilon_j/\varepsilon = \delta_{il}(\varepsilon)v_{en}(\varepsilon)$ is the loss function and $v_j$ is the excitation rate, which is determined at $\varepsilon_{ion} > \varepsilon > \varepsilon_2$ mainly by excitation of the $N_2$ triplet, as well as O($^1$D) and O($^1$S) (Figure 4a). Substituting $St_{cl}(F_e)$ for $ST_{dl}(F_t)$ in equation (5) yields a steady state solution

$$F_e(\varepsilon) = F_t(\varepsilon_2)\exp\left(-\tfrac{3}{2}\int_{\varepsilon_2}^{\varepsilon} d\varepsilon \, \delta_{il}(\varepsilon)/m_e u_\|^2\right) \tag{10}$$

Figure 4 presents the dependence of the key variables on the electric field at various altitudes. The reduced transport frequency in the vibrational barrier decreases with altitude as $\rho_1(h)(1+\gamma_2\rho_2(h)/\rho_1(h))$, where $\gamma_2 = \sigma_2/\sigma_1 \approx \tfrac{1}{3}$ at $\varepsilon \sim$ 2-3 eV. The contribution of $\rho_2(h)(\sigma_D + \sigma_S)$ increases with altitude, while the triplet contribution, $\sim\rho_1(h)\sigma_\Sigma$, decreases. This makes the depth of the local minimum of $\delta_{il}(h)$ at 6-7 eV decrease with $h$ and, similarly, the rise at >10 eV slow down. Figure 4b shows the variation of $v_{vb}\kappa_v(\varepsilon,\tilde{E})$ with $\tilde{E}=\tilde{E}_{vb} \approx 350\rho_1\sqrt{1+\gamma_2\rho_2/\rho_1}$ TD--the upper limit for the "bite-out" approximation. Here, $\tilde{E}_{vb}(h)$ is determined from the violation of condition (8), as illustrated the evolution of the number flux, $\Phi_e(\varepsilon)=2\varepsilon F_e(\varepsilon)/m_e^2$, at $\tilde{E}=\tilde{E}_A$, $\tilde{E}_{ion}$, and $\tilde{E}_{vb}(h)$ (cf. Gurevich, 1978, Figure 8).

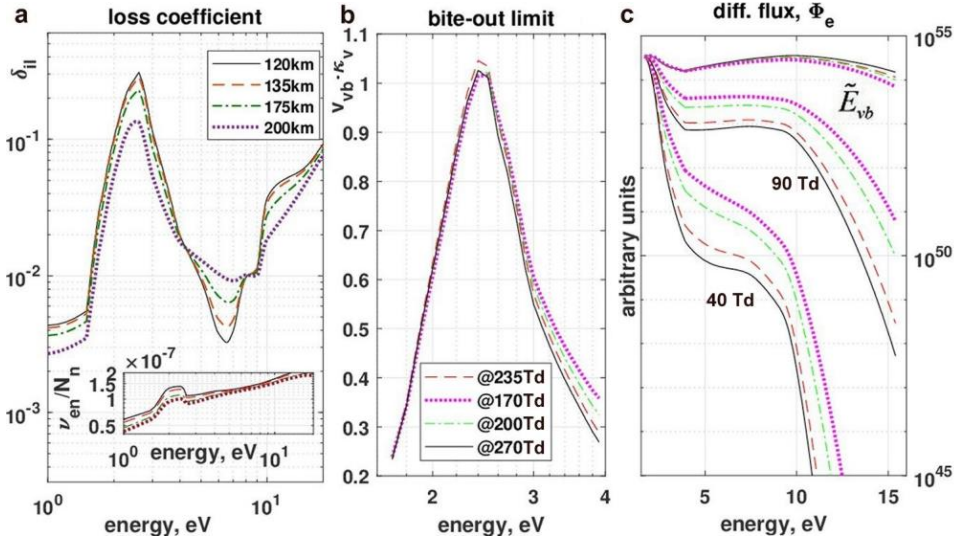

**Figure 4**. The energy dependence at various altitudes of (a) the inelastic loss coefficient, $\delta_{il}(\varepsilon)=v_{il}(\varepsilon)/v_{en}(\varepsilon)$ calculated with the Majeed and Strickland (1997) and Itikawa (2006) data; (b) the product $v_{vb}\kappa_v(\varepsilon,h,\tilde{E}_{vb}(h))$ at the vibrational barrier (see text); and (c) the omnidirectional differential number flux, $\Phi_e(\varepsilon)$, for $\tilde{E}_A$, $\tilde{E}_{ion}$, and $\tilde{E}_{vb}(h)$. Inset: The transport collision frequency, $v_{en}(\varepsilon)/N_n$, per one molecule. The altitudes are indicated by the line style and color, and the electric field magnitudes are shown in Td.

The most important conclusion for the Picket Fence mechanism is that for a broad range of the electric field magnitudes the curves for 120 and 135 km in Figure 4c are close. Accordingly, we can deduce that the energy balance at 135 km remains nearly the same as at 120 km, which was to be demonstrated to explain the Picket Fence color near $h_{pf}$. Notwithstanding that the perturbed atmosphere's upwelling might change the conditions, the estimated altitude range is also consistent with the observations (e.g., Archer et al., 2019b; Liang et al., 2019). Further, in photochemical equilibrium, the volume emission rate (VER), $\eta_\lambda$, is calculated from

$$\eta_\lambda = A_\lambda \cdot [X_\lambda] = A_\lambda \frac{q_\lambda}{L_\lambda + A_{\Sigma^\lambda}} \text{ cm}^{-3}\text{s}^{-1} \qquad (11)$$

Here $[X_\lambda]$ stands for the density of the excited species in cm$^{-3}$; $A_\lambda$, $q_\lambda$, and $L_\lambda$ in s$^{-1}$ are the Einstein transition probabilities, excitation, and loss rates, respectively. For the $B^3\Pi_g$ state, we have $A_{\Sigma^B}=A_B \approx 7.5 \times 10^4 \gg L_B$ yielding $\eta_B \approx q_B$. For O($^1$S) green line ("g") with $A_{\Sigma^g} \approx 1.1 A_g \approx 1.35 \gg L_g$ at $h \geq h_{pf}$, we obtain $\eta_g \approx 0.9 q_g$. For the metastable O($^1$D) state ("r") with

$A_r \approx \frac{7}{9} A_{\Sigma^r} \approx 0.007$ and the collisional quenching rate $L_r(h_{pf}) \geq 1$ s$^{-1}$ (e.g., Mishin et al., 2004, Figure 5), we get $\eta_r \leq 0.007 q_r$. At $\tilde{E} \to \tilde{E}_{ion}$, the flux $\Phi_e(\varepsilon)$ tends to a quasi-plateau between $\varepsilon_2 < \varepsilon \leq \varepsilon_C$ (Figure 4c). The abrupt decrease thereafter is mainly due to the excitation of $C^3\Pi_u$.

Thus, the excitation rates, $q_\lambda \sim N_n \rho_\lambda \int \Phi_e(\varepsilon,h) \sigma_\lambda(\varepsilon) d\varepsilon$, are determined by the area integral $\int \sigma_\lambda(\varepsilon) d\varepsilon$ over the "plateau". Taking that one with the cross-sections in Figure 3a gives $q_r \approx 20 q_g$ and $q_A \approx q_B \approx 2(\rho_1/\rho_2) q_g$, such that $\eta_r \leq 0.16 \eta_g$ and $\eta_B \approx 2(\rho_1/\rho_2) \eta_g$. That is, as anticipated, the green-line and N$_2$1P band emissions dominate the spectrum near $h_{pf}$; the Vegard-Kaplan band via $A^3\Sigma_u^+ \to X^1\Sigma_g^+$ transition are also present. Integrating $\eta_\lambda(h)$ (11) along the line of sight gives the surface brightness

$$I_\lambda = 10^{-6} \int_{h_{min}}^{h_{max}} \eta_\lambda(h)\, dh \quad [R] \tag{12}$$

(1 Rayleigh = $10^6 \frac{\text{photon}}{\text{cm}^2 \text{s}}$). It can be estimated using the power density obtained in Figure 2c, as follows. The average total power consumed near $h_{pf}$ is of the order of $Q_J \sim 0.3$ $\mu$W/m$^3$. A fraction of that, $Q_{tot} \approx (0.1\text{-}0.3) Q_J$, splits between the $N_2$ triplet and O($^1$D) and O($^1$S) states, with each state taking the share $Q_j \approx \varepsilon_j q_j$. Using the obtained relation between the excitation rates yields

$$Q_{tot} \approx \varepsilon_r \left( q_r + q_g \varepsilon_g / \varepsilon_r + q_{A,} \varepsilon_A / \varepsilon_r + q_B \varepsilon_B / \varepsilon_r \right) \approx q_g \varepsilon_r \left[ 22 + 13(\rho_1/\rho_2) \right] \tag{13}$$

Taking $\rho_1(h_{pf}) \sim 2\rho_2(h_{pf}) \sim 2/3$ and the emitting layer of $\Delta h$-km thick ($\Delta h \sim 3$-5), we get $q_g \sim (0.2\text{-}0.6) 10^4$ cm$^{-3}$s$^{-1}$ and the brightness $I_g \sim (0.2\text{-}0.6) \Delta h$ kR, comparable with the IBC1 aurora. The flux in the N$_2$1P band is about four times greater. In general, this value of $I_g$ is the lower limit because of the energy transfer via quenching of the $N_2(A^3\Sigma_u^+)$ state by atomic oxygen

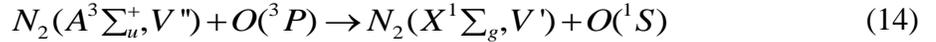

$$N_2(A^3\Sigma_u^+, V'') + O(^3P) \to N_2(X^1\Sigma_g, V') + O(^1S) \tag{14}$$

with the rate coefficient $k_{1S} \approx 2\times 10^{-11}$ cm$^3$/s at $V''=0$ (Piper, 1992). The metastable $A^3\Sigma_u^+$ state has $A_{\Sigma^A} = A_A \approx 0.4$ and $L_A \approx 8\times 10^{-11} \rho_2 N_n \approx 1.5$ s$^{-1}$ at $h_{pf}$. In addition to the direct excitation with $q_A \approx 2(\rho_1/\rho_2) q_g$, the N$_2$1P source, $B^3\Pi_g \to A^3\Sigma_u^+ + h\nu$, complements the rate $q_B \approx q_A$. As a result, equations (11) and (14) yield the O($^1$S) excitation rate twice as the electron impact rate, $q_g$, thereby increasing the green-line brightness, $I_g \sim (0.2\text{-}0.6) \Delta h$ kR, by about three times.

For STEVE well above $h_{pf}$, such comparison with the energy balance in Figure 2c is less reliable, especially for $|E_\parallel| > E_{ion}$. However, note that ionization by the suprathermal electrons will swiftly increase the plasma density at these altitudes. In other words, a strongly depleted density profile in the altitude range ~150-200 km is not sustainable inside strong flow channels. It is, therefore, reasonable to assume that the IFI development in the evolving density profile will be saturated when the generated fields in the whole altitude range reduce to $|E_\parallel| \leq E_{ion}$.

Then, almost the same shape of $\Phi_e(\varepsilon,h)$ as at lower altitudes allows us to draw the following conclusions. While the $N_2$ vibrational excitation still dominates at $\tilde{E} < \tilde{E}_{vb}(h)$, the population permeating the barrier and forming a quasi-plateau increases with altitude. This makes the portion going to the excitation of the $N_2$ triplet and oxygen states increase. The latter, as follows from equation (13), dominates at altitudes where $\rho_1 < 1.7\rho_2$, viz., above ~200 km. Here, the O($^1$D) quenching rate reduces (e.g., Mishin et al., 2004, Figure 5) so that the redline emission dominates the green line and N$_2$1P band, i.e., $\eta_r > \eta_g > \eta_B$. At the same time, transitions between vibrationally excited triplet states facilitate the STEVE continuum. The decrease of the IFI generated field above ~250 km (Figure 2) places the upper limit of the altitude range for STEVE. Similar to Picket Fence, though with the possible atmosphere's upwelling, these predictions are consistent with the observations (e.g., Archer et al., 2019b; Liang et al., 2019).

A final remark is in order. A valley is a common feature of the premidnight subauroral ionosphere (e.g., Titheridge, 2003), which is further deepen inside the SAID channel. Therefore, we suggest that the subauroral arc has the transient phase. Namely, the initial "adjustment" of the density profile features also the $N_2^+$1N blue- and violet-line emissions of the intensity fading out with time. In other words, the initial subauroral spectrum consisting of all typical auroral lines gradually reduces to the discussed above subauroral arc spectrum. Such transition can be revealed in dedicated high-temporal resolution observations.

## 4 Conclusions

We show that the IFI driven by strong electric fields within a low-density trough leads to greatly enhanced $E_\parallel$ and the parallel voltage below the nighttime F$_2$ peak. The presence of the ionospheric valley further increases $E_\parallel$ and the voltage. The obtained electric fields are used as the input into the Boltzmann kinetic equation to find the EDF and the power going to excitation and ionization of neutral species. The simulation results show the feasibility of this mechanism for subauroral arcs. In particular, it creates the population of suprathermal electrons and $N_2$ excitation in good quantitative agreement with that required for Picket Fence. As far as the STEVE spectrum is concerned, the kinetic theory predictions are in a qualitative agreement with its basic features.

### Acknowledgements

E.V. Mishin and A.V. Streltsov acknowledge support by the Air Force Office of Scientific Research LRIR 22RVCOR011 and the NSF award AGS 1803702, respectively.

### Availability Statement

The paper does not produce any new experimental data and numerical codes.